\documentclass{elsart}
\usepackage{graphicx}

\begin{document}
\begin{frontmatter}


\title{Monocular Measurement of the Spectrum of UHE Cosmic Rays by the FADC
Detector of the HiRes Experiment}

\author[Utah]{R.U.~Abbasi}
\author[Utah]{T.~Abu-Zayyad}
\author[LANL]{J.F.~Amman}
\author[Utah]{G.C.~Archbold}
\author[Adelaide]{J.A.~Bellido}
\author[Utah]{K.~Belov}
\author[Montana]{J.W.~Belz}
\author[Rutgers]{D.R.~Bergman\thanksref{email}}
\author[Utah]{Z.~Cao}
\author[Adelaide]{R.W.~Clay}
\author[LANL]{M.D.~Cooper}
\author[Utah]{H.~Dai}                
\author[Adelaide]{B.R.~Dawson}
\author[Utah]{A.A.~Everett}
\author[Utah]{J.H.V.~Girard}
\author[Utah]{R.C.~Gray}
\author[Utah]{W.F.~Hanlon}
\author[LANL]{C.M.~Hoffman}
\author[LANL]{M.H.~Holzscheiter}
\author[Utah]{P.~H\"{u}ntemeyer}
\author[Utah]{B.F Jones}
\author[Utah]{C.C.H.~Jui}
\author[Utah]{D.B.~Kieda}
\author[Utah]{K.~Kim}
\author[Montana]{M.A.~Kirn}             
\author[Utah]{E.C.~Loh}
\author[Tokyo]{N.~Manago}             
\author[LANL]{L.J.~Marek}
\author[Utah]{K.~Martens}
\author[New Mexico]{G.~Martin}
\author[Tokyo]{N.~Manago}
\author[New Mexico]{J.A.J.~Matthews}
\author[Utah]{J.N.~Matthews}
\author[Utah]{J.R.~Meyer}
\author[Utah]{S.A.~Moore}
\author[Utah]{P.~Morrison}
\author[Utah]{A.N.~Moosman}
\author[Utah]{J.R.~Mumford}
\author[Montana]{M.W.~Munro}
\author[LANL]{C.A.~Painter}
\author[Rutgers]{L.~Perera}
\author[Utah]{K.~Reil}
\author[Utah]{R.~Riehle}
\author[New Mexico]{M.~Roberts}
\author[LANL]{J.S.~Sarracino}
\author[Rutgers]{S.~Schnetzer}
\author[Utah]{P.~Shen}
\author[Adelaide]{K.M.~Simpson}
\author[LANL]{G.~Sinnis}
\author[Utah]{J.D.~Smith}
\author[Utah]{P.~Sokolsky}
\author[Columbia]{C.~Song}
\author[Utah]{R.W.~Springer}
\author[Utah]{B.T.~Stokes}
\author[Utah]{S.B.~Thomas}
\author[LANL]{T.N.~Thompson}
\author[Rutgers]{G.B.~Thomson}
\author[LANL]{D.~Tupa}
\author[Columbia]{S.~Westerhoff}
\author[Utah]{L.R.~Wiencke}
\author[Utah]{T.D.~VanderVeen}
\author[Rutgers]{A.~Zech}
\author[Columbia]{X.~Zhang}
                                               
\address[Utah]{University of Utah, Department of Physics and High
Energy Astrophysics Institute, Salt Lake City, Utah, USA}
\address[LANL]{Los Alamos National Laboratory, Los Alamos, NM, USA}
\address[Adelaide]{University of Adelaide, Department of Physics,
Adelaide, South Australia, Australia}
\address[Montana]{University of Montana, Department of Physics and
Astronomy, Missoula, Montana, USA}
\address[Rutgers]{Rutgers - The State University of New Jersey,
Department of Physics and Astronomy, Piscataway, New Jersey, USA}
\address[Columbia]{Columbia University, Department of Physics and
Nevis Laboratory, New York, New York, USA}
\address[New Mexico]{University of New Mexico, Department of Physics
and Astronomy, Albuquerque, New Mexico, USA}
\address[Tokyo]{University of Tokyo, Institute for Cosmic Ray
Research, Kashiwa, Japan}
\collaboration{The High Resolution Fly's Eye Collaboration}

\thanks[email]{Corresponding author, E-mail:
\texttt{bergman@physics.rutgers.edu}}

\date{\today}

\begin{abstract}
  We have measured the spectrum of UHE cosmic rays using the Flash ADC
  (FADC) detector (called HiRes-II) of the High Resolution Fly's Eye
  experiment running in monocular mode.  We describe in detail the
  data analysis, development of the Monte Carlo simulation program,
  and results.  We also describe the results of the HiRes-I detector.
  We present our measured spectra and compare them with a model
  incorporating galactic and extragalactic cosmic rays.  Our combined
  spectra provide strong evidence for the existence of the spectral
  feature known as the ``ankle.''
\end{abstract}

\end{frontmatter}




\section{Introduction}

The aim of the High Resolution Fly's Eye (HiRes) experiment is to
study the highest energy cosmic rays using the atmospheric
fluorescence technique.  In this paper we describe the data
collection, analysis, and Monte Carlo calculations used to measure the
cosmic ray spectrum with the HiRes experiment's FADC detector,
HiRes-II.  We also describe the analysis performed on the data
collected by the HiRes-I detector and present the two monocular
spectra, covering an energy range from $2 \times 10^{17}$ eV to over
$10^{20}$ eV.  We perform a statistical test of the combined spectra
which gives strong evidence for the presence of the spectral feature
known as the ``ankle.''  We conclude with a fit of our data to a toy
model incorporating galactic and extragalactic cosmic ray sources.

The acceleration of cosmic rays to ultra high energies is thought to
occur in large regions of high magnetic fields expanding at
relativistic velocities\cite{kn:acceleration}.  Such structures
are rare in the neighborhood of the Milky Way galaxy and many of the
cosmic rays that we observe may have traveled cosmological distances
to reach us.  Hence they are probes of conditions in some of the most
violent and interesting objects in the universe.

The highest energy particles from terrestrial particle accelerators
have energy $1 \times 10^{12}$ eV, so the cosmic rays we observe have
energies at least five orders of magnitude higher.  Since we observe
showers in the atmosphere initiated by the cosmic ray particles, we
are sensitive to their composition and to the details of their
interactions with matter. 

Interactions of high energy protons, traveling large distances across
the universe, with photons of the cosmic microwave background
radiation can excite nucleon resonances which decay to a nucleon plus
a $\pi$ meson.  This is an important energy loss mechanism for the
cosmic rays, and results in the Greisen-Zatsepin-Kuzmin (GZK)
cutoff\cite{kn:gzk}, which is often stated as: cosmic rays traveling
more than 50 Mpc should have a maximum energy of $6 \times 10^{19}$
eV, if sources are uniformly distributed.  Several events above this
energy have been seen by previous
experiments\cite{kn:prevexp,kn:flyseye,kn:agasa}, but statistics are
low and it is crucial to search for more events above the GZK cutoff.

The spectrum of cosmic rays has few distinguishing features.  It
consists of regions of power law behavior with breaks in the power law
index.  There is a steepening from E$^{-2.7}$ to E$^{-3.0}$ at about
$3 \times 10^{15}$ eV (called the knee)\cite{kn:knee} and a hardening
at higher energy (called the ankle).  The Fly's Eye
experiment\cite{kn:flyseye}, observing in stereo mode, saw a second
knee (or steepening of the spectrum) at $4 \times 10^{17}$ eV and the
ankle at $3 \times 10^{18}$ eV.  The second knee has also been
observed by the Akeno experiment\cite{kn:akeno}.  The Haverah Park
experiment\cite{kn:hpark} observed the ankle at about $4 \times
10^{18}$ eV. The Yakutsk experiment\cite{kn:yakutsk} has seen both the
second knee and the ankle.  The AGASA experiment\cite{kn:agasa}, which
has a large enough aperture to collect events with energies of
$10^{20}$ eV, observes a higher flux than Fly's Eye, and the ankle at $1
\times 10^{19}$ eV.  They observe a dip at the GZK threshold, but
their spectrum then recovers at higher energies.

The atmospheric fluorescence technique has its basis in the fact that,
on average, approximately five UV fluorescence
photons\cite{kn:kakimoto} will be emitted when a minimum ionizing
particle of charge $e$ passes through one meter of air.  In HiRes, we
detect these photons and reconstruct the development of cosmic ray air
showers.  We collect the fluorescence light with spherical mirrors of
area 5.1 m$^2$, and focus it on a $16 \times 16$ array of
photomultiplier tubes, each of which looks at about one degree of the
sky.  We record the integrated pulse height and trigger time
information from each tube, and can reconstruct the geometry of the
air shower and the energy of the primary cosmic ray that initiated it.

HiRes consists of two detector sites located on desert hilltops on the
U. S. Army's Dugway Proving Ground in west central Utah.  The first
site, called HiRes-I, consists of 22 detectors that look between 3 and
17 degrees in elevation and almost 360 degrees in azimuthal
angle\cite{kn:HiRes-INIM}.  This detector uses an integrating ADC
readout system which records the photomultiplier tubes' pulse height
and time information.

The second site, called HiRes-II and located 12.6 km away, consists of
42 detectors looking between 3 and 31 degrees in elevation, and has a
Flash ADC (FADC) system to save pulse height and time information from
its phototubes\cite{kn:HiRes-IINIM}.  The sampling period of the FADC
electronics is 100 ns.  Cosmic ray air showers with energies near
$10^{20}$ eV and occurring within a radius of 35 km, can trigger the
HiRes detectors and can be reliably reconstructed.

The two detector sites are designed to observe cosmic ray showers
steroscopically.  This stereo mode observation gives us the best
geometric resolution, about 0.6 degrees in pointing angle and 100 m in
distance to the shower.  In this mode we make two measurements of the
particle's energy and thus can make an empirical determination of our
energy resolution.  The limitation of stereo mode is a geometrically
imposed lower energy threshold of $10^{18}$ eV.  At this energy the
events lie halfway between the two detectors, about 6 km from each.

In monocular mode, the HiRes-II detector can observe events much
closer and dimmer than is possible in stereo mode; the energy
threshold for this mode is about $2 \times 10^{17}$ eV.  The geometric
resolution is still good: about 5 degrees in pointing angle and 300 m
in distance.  In this paper we describe the operation and data
analysis for the HiRes-II detector, briefly describe the differences
between HiRes-I and HiRes-II, and present the monocular spectra of the
two detectors.

\section{Calibration Issues}

There are two important calibration issues in HiRes: the first is the
absolute calibration of the phototubes' pulse heights in photons.
This is accomplished by carrying a standard light source to each of
our detectors and illuminating the phototubes with
it\cite{kn:abscalib}.  This source is absolutely calibrated using NIST
calibrated photodiodes to about 10\% accuracy and this uncertainty
appears in our energy measurements.

Since the atmosphere is both our calorimeter and the medium through
which we look, we must correct for the way it absorbs and scatters
fluorescence light.  The determination of the characteristics of the
atmosphere is our second important calibration.  Both the molecular
and aerosol components of the atmosphere contribute to the scattering.
The molecular component is well known, but we must measure the aerosol
component's contribution.  Two steerable YAG lasers, one at each site
and operating at wavelength $\lambda=355$ nm, are used for the aerosol
calibration.  The scattered light from the laser at one detector is
observed by the other detector.  In this way we measure the scattering
length, angular distribution of the scattering cross section, and
vertical aerosol optical depth (VAOD) of aerosol particles in the
atmosphere\cite{kn:avgatm}.  The aerosol scale height is obtained from
the product of horizontal extinction length at ground level times the
VAOD.

\begin{figure}[tbh]
  \includegraphics[width=0.67\columnwidth]{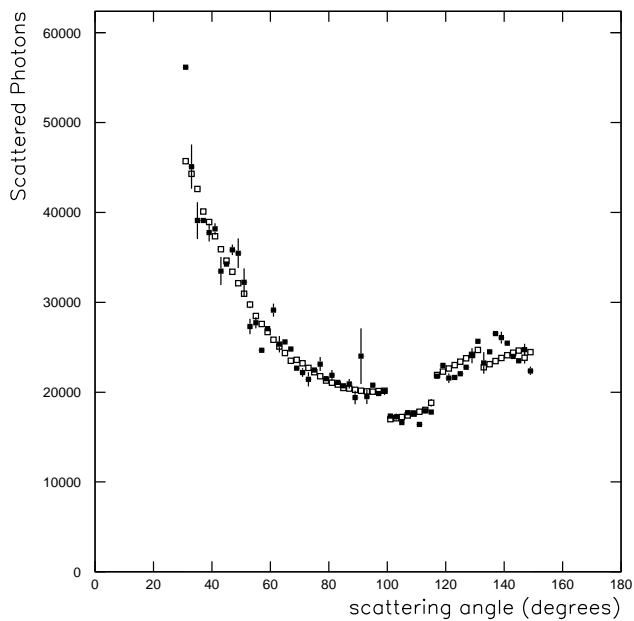}
  \caption{Intensity of laser light scattered into the HiRes-I
    detector plotted against the scattering angle of the light.  The
    filled squares are the data and the open squares are a
    four-parameter fit to the data.  The discontinuity between 100 and
    116 degrees is due to the parallax of a mirror farther from the
    laser track.}
  \label{fig:36deg}
\end{figure}

Figure \ref{fig:36deg} shows the amount of light detected as a
function of scattering angle for one of these laser events.  This shot
was fired horizontally from the HiRes-II site and passed within 400
meters of the HiRes-I detector.  This geometry allows us to observe a
wide range of scattering angles.  The filled squares show the data,
and the open squares are a fit to this data using a four parameter
model of the aerosol extinction length and angular distribution.  The
forward peak seen in this figure is characteristic of aerosol
scattering and the relatively flat distribution at backward scattering
angles is characteristic of molecular scattering.  The horizontal
aerosol extinction length measured from this laser event is 23 km.
For comparison, the horizontal molecular extinction length at this
same wavelength (355 nm) is 18 km.  These extinction lengths
correspond approximately to the average of atmospheric conditions
during our observations at Dugway.

We perform the laser measurement of atmospheric conditions hourly
during data collection.  For the analysis reported here the average of
hourly aerosol scattering lengths and scale heights were used.  Since
the data has good statistics and the events were collected evenly over
the period in question, they will be well described by the average
atmospheric conditions \cite{kn:avgatm}, which were: aerosol
scattering length of $22 \pm 2$ km and scale height of 1.1 km.  The
RMS of the scale height distribution was 0.4 km, and the systematic
uncertainty was smaller than this.

\section{Data Analysis}

The FADC data acquisition system records a 10 $\mu$s long series of
ADC samples (100 samples total) for each active photomultiplier tube
(PMT) in an event.  The starting time of the series is chosen to have
the peak of the signal pulse in the middle of the sample.

\begin{figure}[tbh]
  \includegraphics[width=\columnwidth]{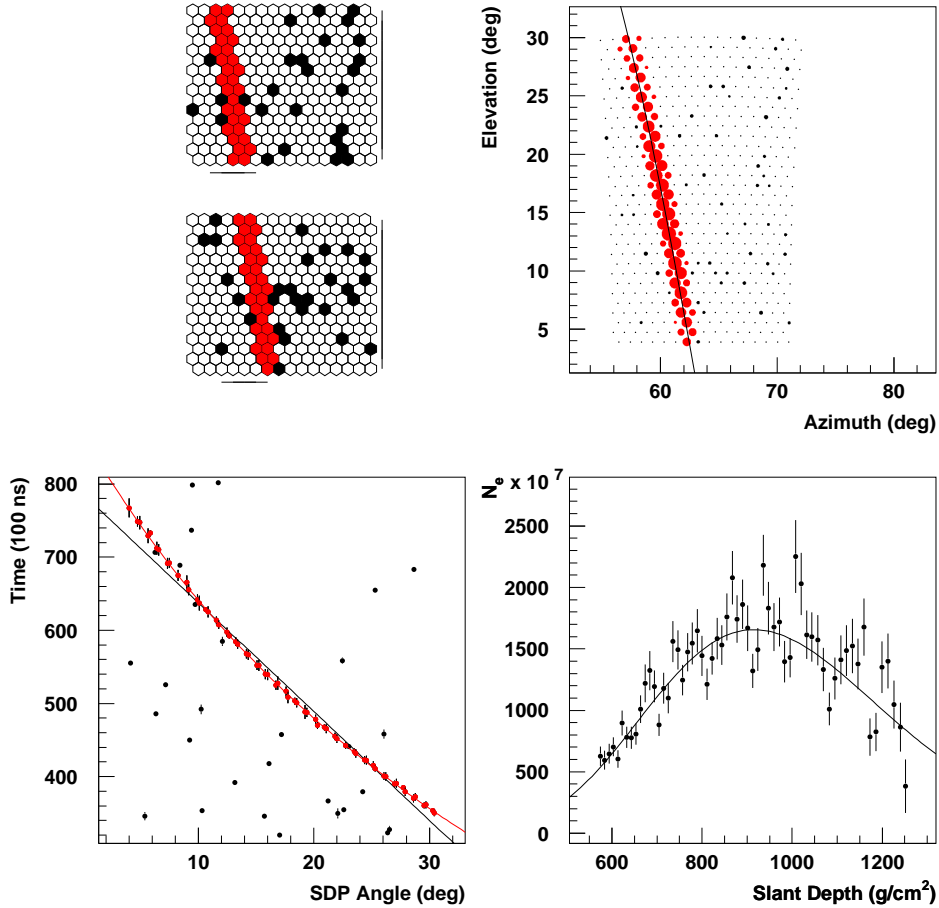}
  \caption{Display of an Event with a Reconstructed Energy of
    $2.4\times10^{19}$ eV.  The upper left part of this figure shows
    the two mirrors that triggered for this event. The upper right
    panel shows the azimuthal vs. elevation angles of triggered tubes,
    with a fitted shower-detector plane superimposed.  The lower left
    panel shows the time of the tube hits in FADC time slices vs. the
    angle of the tube measured along the track, with two fits
    superimposed: a straight line and the result of the time fit.  The
    lower right quarter shows the number of charged particles in the
    shower as a function of slant depth (in g/cm$^2$), with the fit to
    the Gaisser-Hillas formula (Eq~\ref{eq:GH}) superimposed.}
  \label{fig:event}
\end{figure}

The first step in the analysis of the data consists of pattern
recognition to choose which hit tubes were on the track of the cosmic
ray event.  As a cosmic ray shower propagates down through the
atmosphere, the mirrors collect the generated photons and focus them
onto the arrays of phototubes.  The image moves across the array
illuminating one or more tubes at a time.  Therefore, tubes on the
cosmic ray track are near each other in two ways: spatially and
temporally.  Phototubes must be near each other in both position and
time to be included in the track.  The top two quarters of Figure
\ref{fig:event} show the picture of an event, where one can see that
the tubes on the shower form a line.  The lower left part of this
figure is a time plot: a plot of the light arrival times (in FADC
time-bin units) on the vertical axis versus the angle of the tube
measured along the track.  From these plots, it is clear that the
tubes related to the air shower can be separated from those firing
from random sky fluctuations.  The elevation and azimuthal angles of
the PMT's on the track are fit to determine the plane which contains
the shower and the detector.

\begin{figure}
  \includegraphics[width=0.67\columnwidth]{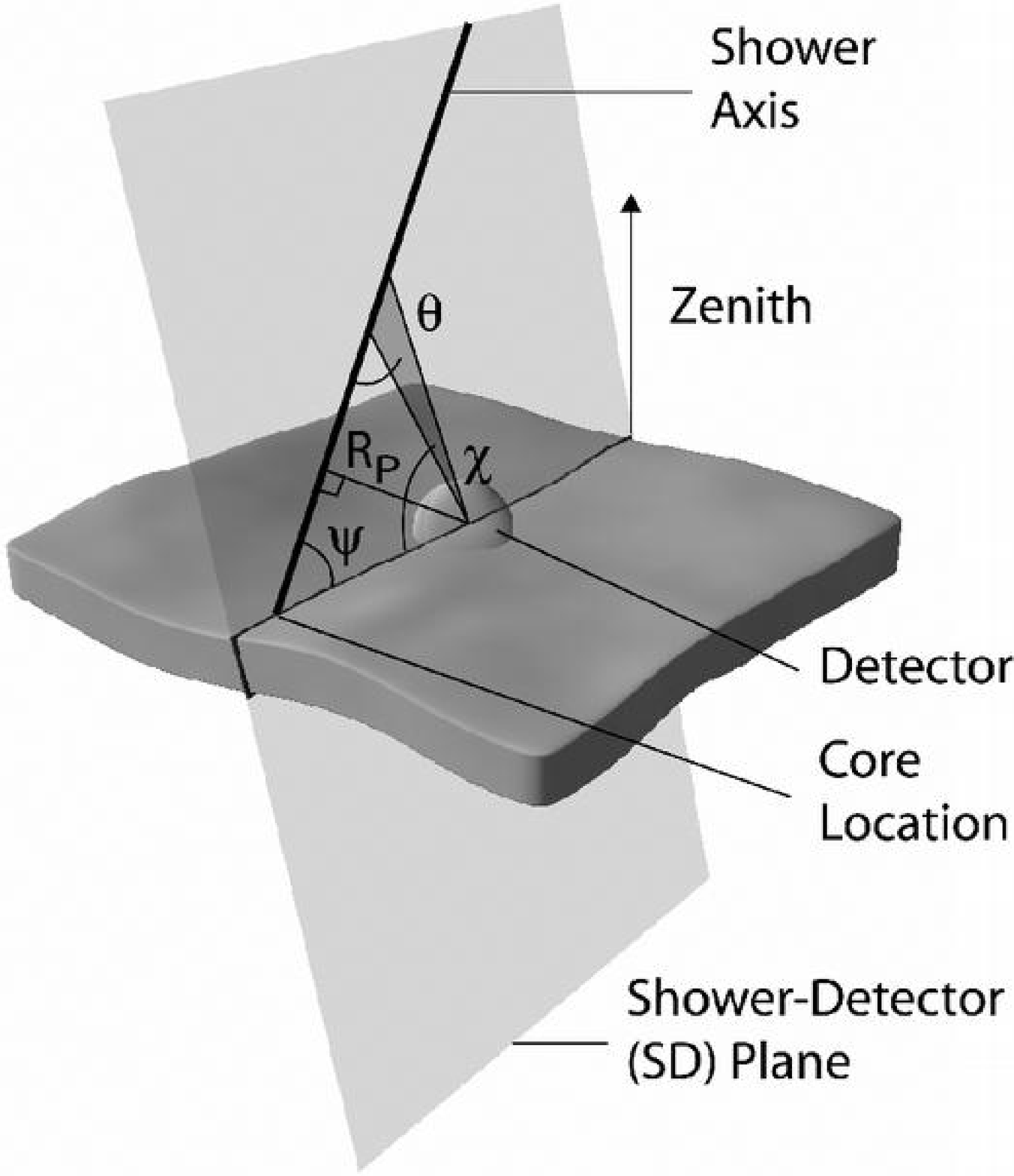}
  \caption{Illustration of monocular reconstruction of the shower
    geometry, showing the shower-detector plane, the impact parameter,
    $R_p$, and the in-plane angle shower angle, $\psi$.}
  \label{sdplane}
\end{figure}

In a monocular determination of the shower geometry, the angle of the
shower within the shower detector plane is determined from the time
plot of the active tubes (see the lower left quadrant of Figure
\ref{fig:event}).  One can show that
\begin{equation}
        t_i = t_0 + \frac{R_p}{c}
                        \tan\left(\frac{\pi-\psi-\chi_i}{2}\right)
\label{eq:tvsa}
\end{equation}
where $t_i$ is the arrival time of light from shower segment $i$,
$\chi_i$ is the angle in the plane containing the shower and detector
from the ground to segment $i$, $t_0$ is the earliest possible arrival
time, $R_p$ is the impact parameter of the shower, and $\psi$ is the
angle the shower makes with the ground in the shower-detector plane.
The geometry of the shower detector plane is shown in
Figure~\ref{sdplane}.  We measure $t_i$ and $\chi_i$, and need to fit
for $t_0$, $\psi$ and $R_p$.  $\psi$ and $R_p$ determine the geometry
within the shower-detector plane.  Since equation \ref{eq:tvsa} is
linear in $R_p$ and $t_0$, we fit for those variables for fixed values
of $\psi$ from $5^\circ$ to $75^\circ$ in $1^\circ$ steps.  The best
geometry is chosen by minimizing $\chi^2$, and the uncertainty in
$\psi$ from the angles which increase the $\chi^2$ by one.

Once the geometry of the shower is known, we can reconstruct the
number of charged particles in the shower as a function of the slant
depth of atmosphere through which the shower has passed.  To do this,
we collect the photoelectrons from tubes on the track into successive
time bins, which are multiples of the FADC sampling period.  Typically
several tubes will contribute to each time bin.  Systematic errors in
calculating the acceptance of individual tubes tend to be offset by
correlated errors in neighboring tubes, reducing the overall
uncertainty in the acceptance calculation.  We then correct for the
sum of the acceptance of all the participating PMT's and for the
quantum efficiency of the phototubes, the mirror reflectivity and the
transmission of the HiRes UV filter.  This yields the flux of photons
striking the mirror.

To convert the photon flux at the detector into the number of charged
particles at the observed position of the shower\cite{kn:balt}, we
first correct for the solid angle of the mirror with respect to the
shower.  We then correct for the amount of light lost due to
scattering and absorption in the atmosphere.  This includes light
scattered by Raleigh scattering from air molecules, Mie scattering
from aerosol particles and absorption due to ozone.  The first
calculation of the attenuation correction is done assuming that all
the observed photons come from the fluorescence spectrum given in
Bunner\cite{kn:bunner}.  The solid angle and attenuation corrections
give the photon flux at the observed portion of the shower.  Finally,
we calculate the charged particle multiplicity at the shower using the
fluorescence yield measurements of Kakimoto {\it et
  al}\cite{kn:kakimoto}.

The charged particle multiplicity distribution is fit to the
Gaisser-Hillas profile function\cite{kn:ghfit}:
\begin{equation}
N(X)=N_{max}(\frac{X-X_0}{X_{max}-X_0})^{\frac{X_{max}-X_0}{\lambda}}
\exp(\frac{X_{max}-X}{\lambda}),
\label{eq:GH}
\end{equation}
where $N(X)$ is the number of charged particles in the shower at slant
depth $X$, $N_{max}$ is the number of particles at shower maximum,
$X_{max}$ is the slant depth of the maximum, and $\lambda$ is a
shower-development parameter.  We have seen in previous measurements
that the Gaisser-Hillas profile function fits extensive air showers
very well\cite{kn:caozh}.  Our fits are very insensitive to $X_0$ and
$\lambda$ and we fix them at -60 and 70 g/cm$^2$, respectively.  The
$X_0$ value is chosen to agree with our fits to Corsika showers (see
below).  

We calculate the correction for scattered \v{C}erenkov light as
follows.  We use the fitted Gaisser-Hillas function to simulate the
development of the beam of \v{C}erenkov photons accompanying the
shower, and calculate the number of these photons scattered into our
detector acceptance.  The atmospheric attenuation is recalculated for
the mixture of fluorescence and scattered \v{C}erenkov photons, and
the relative numbers of photoelectrons from fluorescence and
\v{C}erenkov sources are found after applying the filter transmission
and quantum efficiency corrections using the appropriate spectra.  The
photoelectrons from \v{C}erenkov photons are subtracted from the
signal, and the charged particle multiplicity is recalculated again as
described above.  This iterative process is continued until stability
is achieved.  This \v{C}erenkov correction is typically about 15\%.
The lower right part of Figure \ref{fig:event} shows the development
profile of a shower after the correction has been performed, and the
Gaisser-Hillas function fit to this profile.

We integrate the final fitted Gaisser-Hillas function over all $X$ and
multiply by the average energy loss per particle (2.19 MeV/g/cm$^2$)
to determine the visible shower energy.  The visible energy is then
corrected for energy carried off by unobservable
particles\cite{kn:chiwha} to give the total shower energy.

Cuts are applied to select well-reconstructed events and to assure
good resolution.  The cuts used in the determination of the UHE cosmic
ray spectrum are listed below:
\begin{itemize}
\item Angular speed $< 11^\circ\ \mu s^{-1}$
\item Selected tubes $\ge 7$
\item 0.85 $<$ Tubes/degree $<$ 3.0
\item Photoelectrons/degree $>$ 25
\item Track length $> 7^\circ$, or $> 10^\circ$ for events extending
above $17^\circ$ elevation
\item Zenith angle $< 60^\circ$
\item 150 $< X_{max} < 1200 {\rm\ g/cm}^2$, and is visible in detector
\item Average \v{C}erenkov Correction $<$ 60\%
\item Geometry fit $\chi^2$/d.o.f. $<$ 10
\item Profile fit $\chi^2$/d.o.f. $<$ 10
\end{itemize}

\section{Development of the Monte Carlo Simulation Program}

We calculated the aperture of the detector using two Monte Carlo
simulation programs.  First we generated a library of cosmic ray
showers using the programs CORSIKA\cite{kn:corsika} and
QGSJET\cite{kn:qgsjet}.  We then use events from the library as input
to a second program which calculates the response of the detector and
writes out simulated events in the same format as the data.  Finally,
we analyze the Monte Carlo events using the same programs used for the
data.

The shower library consists of 200 showers with proton and 200 showers
with iron primaries generated for each combination of five fixed
primary energies from $10^{16}$ eV to $10^{20}$ eV and three fixed
zenith angles of the shower axis with a secant of 1.00, 1.25 and 1.50.
Each shower is characterized by its depth of first interaction in the
atmosphere, energy, zenith angle, type of primary particle, and the
four parameters of a Gaisser-Hillas fit to its profile (the
Gaisser-Hillas formula fits CORSIKA + QGSJET showers very well).

\begin{figure}[tbh]
  \includegraphics[width=0.67\columnwidth]{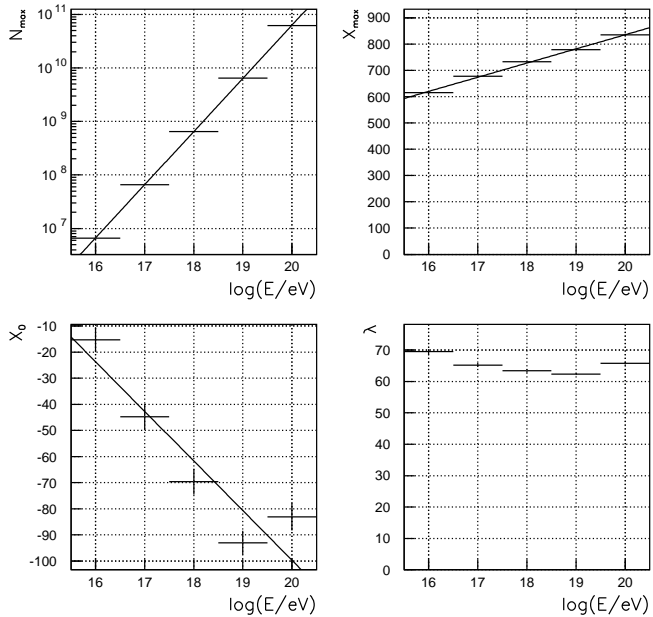}
  \caption{Energy Dependence of Gaisser-Hillas Parameters.  The four
    panels show (clockwise from upper left) the energy dependence of
    $N_{max}, X_{max}, \lambda$, and $X_0$ for showers at zenith angle
    of 36.9 degrees.}
  \label{fig:g-h}
\end{figure}

When we use these events, we must scale their parameters in energy
from the (discrete) energies of the shower library to the continuous
energy spectrum we throw in the detector-response Monte Carlo program.
Figure \ref{fig:g-h} shows the energy dependence of the four
Gaisser-Hillas parameters.  In scaling the parameters of a shower we
use the slopes shown in the four parts of this figure.  Use of a
shower library preserves the event-to-event fluctuations and
correlations in the CORSIKA events.

Since we change the geometry of the showers at random, one CORSIKA
shower can be used over and over to create different events. This
allows us to generate approximately 30 times as many events per minute
with the shower library as we could directly with CORSIKA.

The detector-response program simulates the generation of fluorescence
and \v{C}erenkov light by the shower and the operation of the two
HiRes detectors, including optics, trigger, electronics, and data
acquisition. To generate an event, the program chooses the primary
energy and the primary particle type from the spectrum and composition
measured in stereoscopic mode by the Fly's Eye
experiment\cite{kn:flyseye}.  The zenith angle and distance to the
shower are chosen randomly.  An event from the shower library bin
whose fixed energy and zenith angle are closest to the chosen values
is then used to generate the profile of the shower's development.  We
scale each of the four Gaisser-Hillas parameters to the thrown energy.
The dependence of the Gaisser-Hillas parameters on zenith angle is
quite weak, hence we simply use the three bins in zenith angle.

An accurate simulation of fluorescence and \v{C}erenkov light is
performed~\cite{kn:chiwha}, including the shower profile, the average
$dE/dx$ for each part of the shower, and atmospheric pressure, the
width of the showers, the energy of particles that fall below the
Corsika thresholds (we use 0.1 MeV for electrons and photons, 0.3 GeV
for hadrons, and 0.7 GeV for muons), calorimetric energy, and the
unobserved energy (mostly neutrinos and muons that strike the ground).

Previous publications describe how we calculate fluorescence and
\v{C}erenkov light emission, scattering, and
transmission\cite{kn:balt}.  The fluorescence spectrum is taken from
Bunner {\it et al.}\cite{kn:bunner}, and the overall normalization
from Kakimoto {\it et al}\cite{kn:kakimoto}.  The response due to
mirror reflectivity, HiRes filter transmission, and phototube quantum
efficiency is included.  A complete wavelength-dependent calculation
is performed for all these effects in 16 wavelength bins between 290
and 410 nm.

To simulate the exact conditions of the experiment, we created a
database of parameters that vary from night to night: live time,
trigger logic, trigger gains and thresholds, and specific mirrors in
operation.  Two parameters which vary with time, but which we treated
only in an average way, are the sky noise and atmospheric scattering
of fluorescence light.

These parameters are read into the detector response programs
individually for each event, allowing us to simulate precisely the
detector settings recorded during data collection.  Direct comparisons
of Monte Carlo events and real data, such as those shown below, give
us confidence in our detector response programs and prove that we
understand our detectors.

\begin{figure}[tbh]
  \includegraphics[width=0.67\columnwidth]{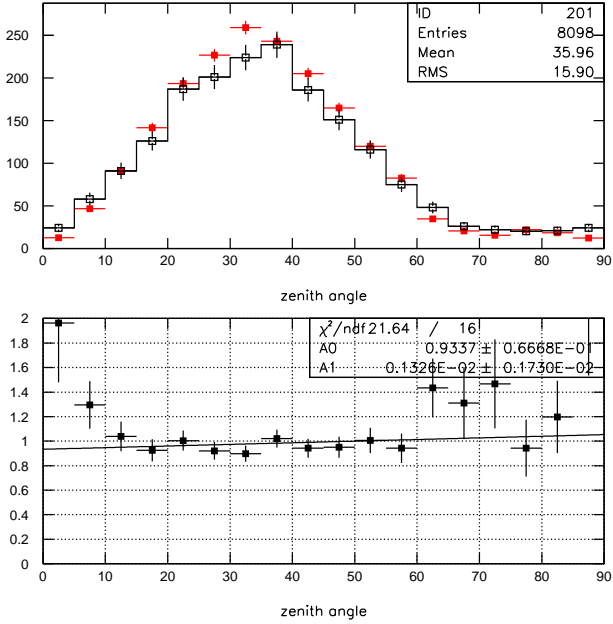}
  \caption{Zenith Angle of Cosmic Ray Showers.  In the upper panel the
    data is shown as open squares and histogram and the Monte Carlo
    (which has been normalized to the number of data events) is shown
    as closed squares.  The lower panel shows the ratio of data to
    Monte Carlo events.}
  \label{fig:zenang}
\end{figure}

\begin{figure}[tbh]
  \includegraphics[width=0.67\columnwidth]{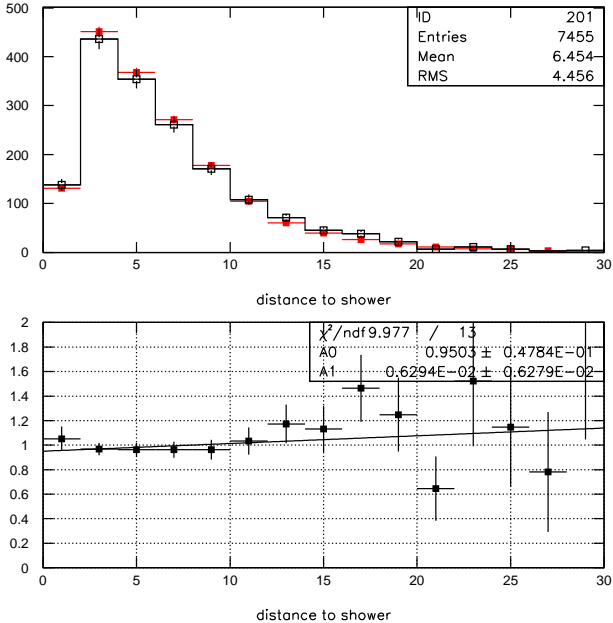}
  \caption{Distance from the Detector to the Shower Mean (weighted by
    photoelectrons).  Again data is shown as open squares and
    histogram and Monte Carlo as closed squares, and the lower panel
    shows the data to Monte Carlo ratio.}
  \label{fig:distmean}
\end{figure}

\begin{figure}[tbh]
  \includegraphics[width=0.67\columnwidth]{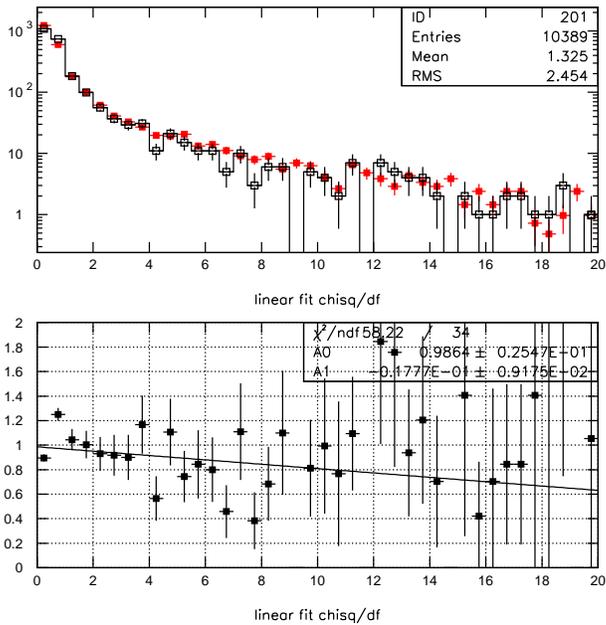}
  \caption{Comparison of data and MC for the $\chi^2$ of a linear fit
    to the time vs angle plot.}
  \label{fig:chisq}
\end{figure}

The data that went into the comparison plots shown below were recorded
by the HiRes-II-detector from 1 December 1999, through 4 May 2000.
There are about 2100 events after cuts. The Monte Carlo sample
contains about five times as many events. The first two graphs
presented here (see Figures~\ref{fig:zenang} and \ref{fig:distmean})
show two basic geometric quantities: the zenith angle distribution and
the distance to the shower mean (found by weighting each PMT that was
on the track by the number of observed photoelectrons). The upper
panels of the graphs show the data as open squares and histograms and
the Monte Carlo as filled squares.  The data and MC distributions have
been normalized to the same area.  In the lower panels, the ratio of
data divided by MC and a linear fit to this ratio are shown.  It can
be seen from Figures \ref{fig:zenang} and \ref{fig:distmean} that the
distributions of these geometric quantities agree very well.  Figure
\ref{fig:chisq} shows the $\chi^2$ of a linear fit to the time plot
(such as is shown in the lower left quadrant of Figure
\ref{fig:event}).  The agreement shows that the experimental
resolution is well simulated in the Monte Carlo program.  

\begin{figure}[tbh]
  \includegraphics[width=0.67\columnwidth]{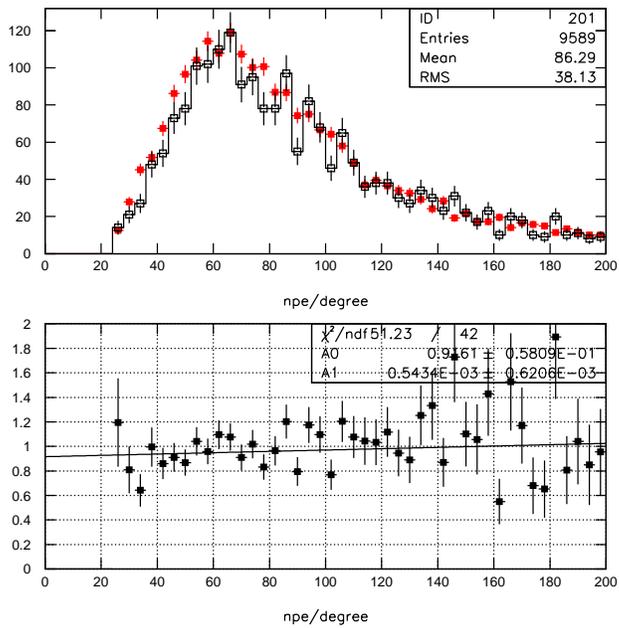}
  \caption{Comparison of data and MC for the photoelectrons per degree
    of track}
  \label{fig:npe}
\end{figure}

\begin{figure}[tbh]
  \includegraphics[width=0.67\columnwidth]{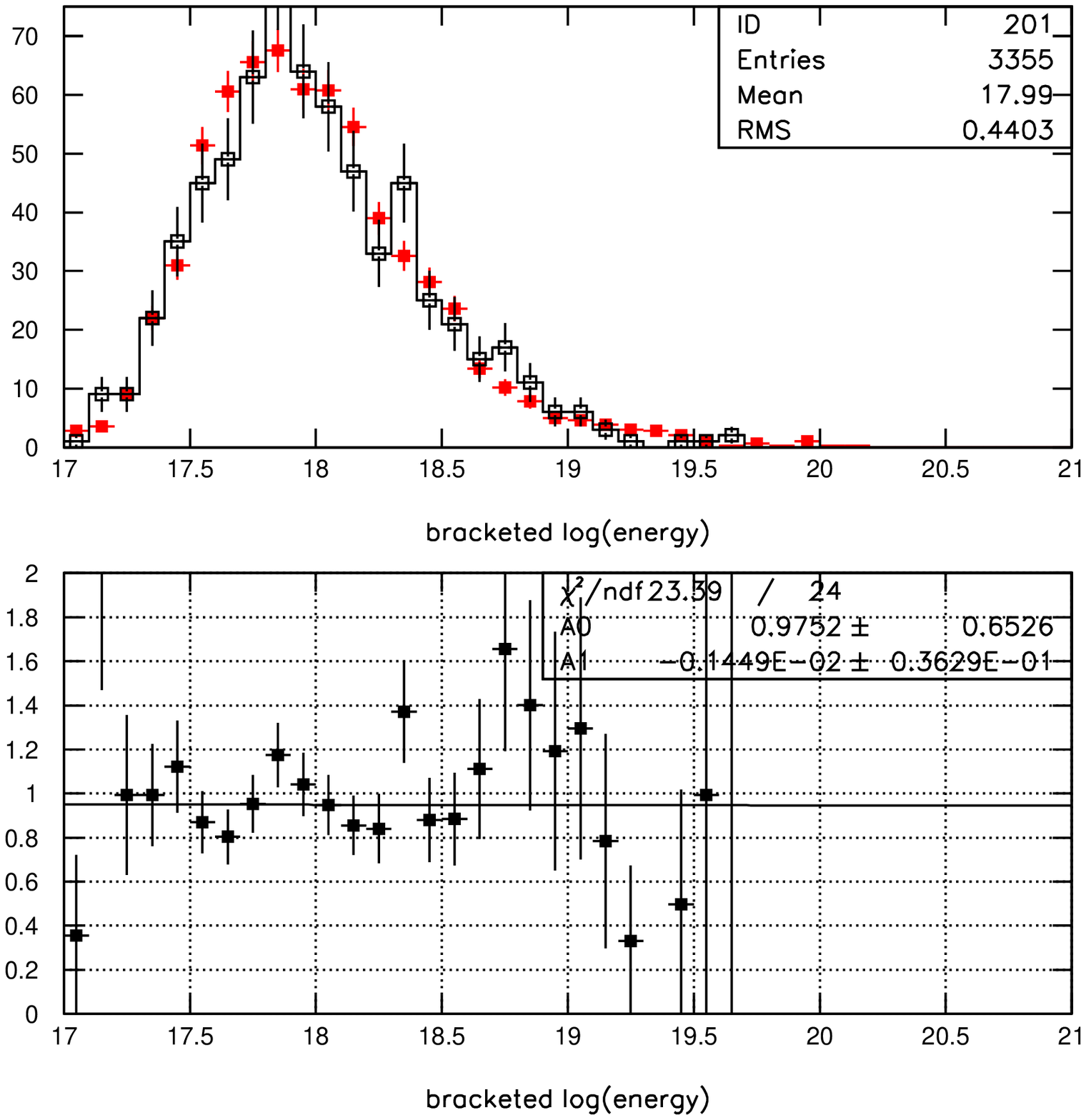}
  \caption{Comparison of data and MC for the reconstructed Energy}
  \label{fig:energy}
\end{figure}

An important non-geometric quantity is the amount of light that is
seen by the detector. It can be characterized by the number of
photoelectrons we receive per degree of track length. Figure
\ref{fig:npe} shows that the amount of light we see with our detectors
and the amount of light we generate in our MC programs closely agree
with each other.  Figure \ref{fig:energy} shows a histogram of the
reconstructed energy of events.

The excellent agreement between the data and Monte Carlo simulation in
these plots is characteristic of our Monte Carlo as a whole and
demonstrates that the Monte Carlo models the data well.

\section{The UHE Cosmic Ray Spectrum}

\begin{figure}[tbh]
  \includegraphics[width=0.67\columnwidth]{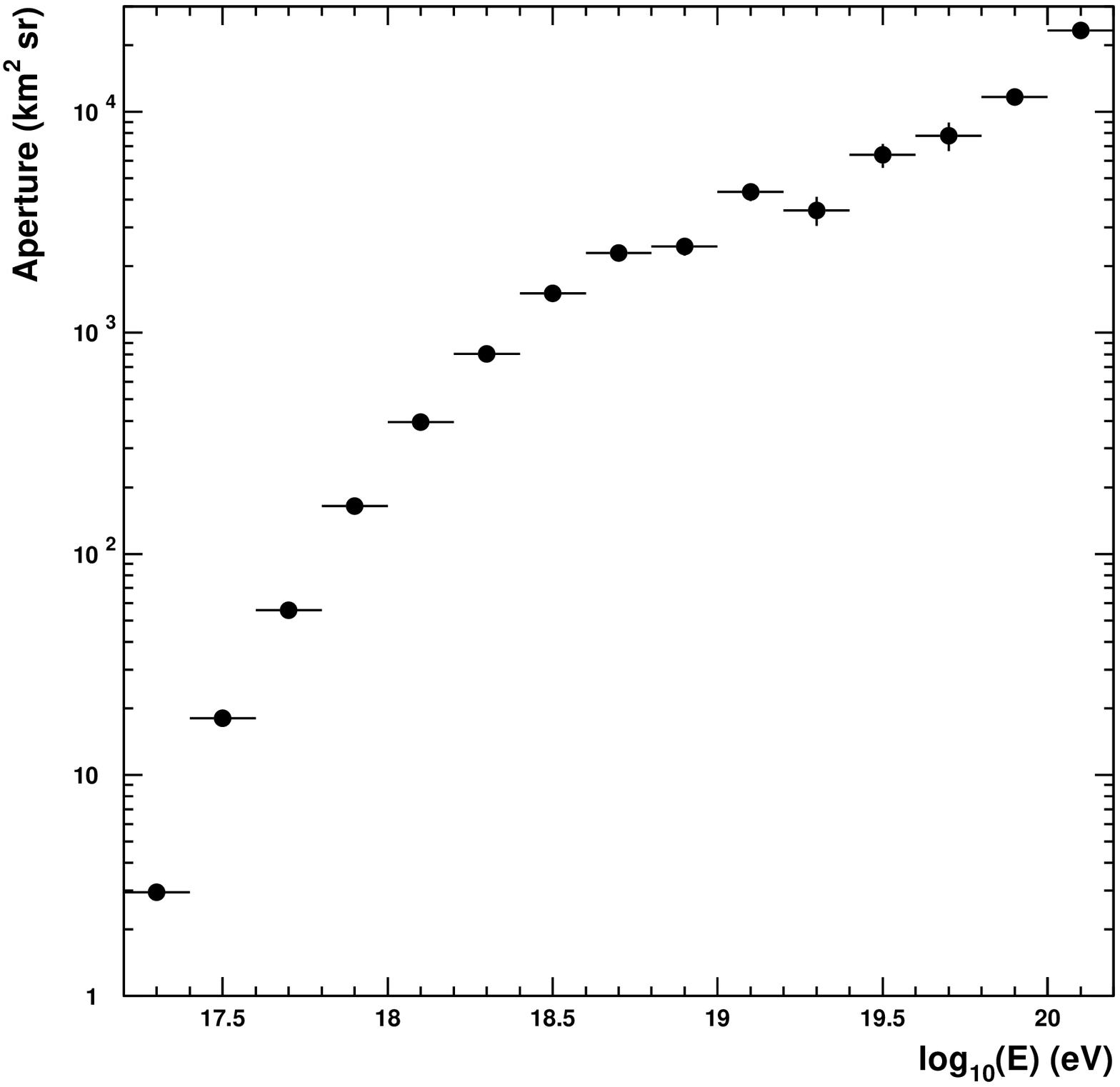}
  \caption{Monte Carlo calculated aperture of the HiRes-II detector
    used for this spectrum measurement.}
  \label{fig:aperture}
\end{figure}

Having demonstrated that our MC models the detector accurately, we
have confidence in using it to calculate the detector aperture.  This
aperture is shown in Figure~\ref{fig:aperture}.

To make an accurate calculation of the flux of cosmic rays it is
important to use a continuous Monte Carlo input spectrum in order to
take account of the finite energy resolution of the detector.  With
this in mind, we define the flux, $J(E)$, as follows:
\begin{equation}
        J(E) = N_D(E)
                        \frac{N_T(E)}
                             {N_A(E)}
                \frac{1}{\Delta E A \Omega T}
\label{eq:jofe}
\end{equation}
where $N_D(E)$ is the number of data events in energy bin $E$,
$N_T(E)$ is the number of thrown MC events in energy bin $E$ binned by
the thrown energy, $N_A(E)$ is the number of accepted MC events in
energy bin $E$ binned by the reconstructed energy, $\Delta E$ is the
width of energy bin $E$, $A$ is the area into which the MC generated
events, $\Omega$ is the solid angle into which the MC generated
events, and $T$ is the total running time of the detector.  The MC
generated events within a 35 km radius of the detector and with zenith
angles from 0$^\circ$ to $70^\circ$.  For the data included in this
paper, recorded from 1 December 1999 to 4 May 2000, the detector was
live for 144 hours.  This includes data only from nights with good
weather.  This time period represents the first period of stable
running for the HiRes-II detector.  After this period the trigger was
changed considerably, so subsequent data have to be analyzed
separately.

An important feature of Equation \ref{eq:jofe} is that when one has
modeled the experimental resolution correctly and put in the correct
thrown energy spectrum, $N_T(E)$, the ratio $N_D(E)/N_A(E)$ becomes a
constant independent of energy.  In this situation, one makes a first
order correction for experimental resolution\cite{cowan}; the spectrum
one calculates has the shape of $N_T(E)$.  The comparisons between
data and Monte Carlo (see especially Figures \ref{fig:chisq} and
\ref{fig:energy}) show that our modeling is accurate.

\begin{figure}[tbh]
  \includegraphics[width=0.67\columnwidth]{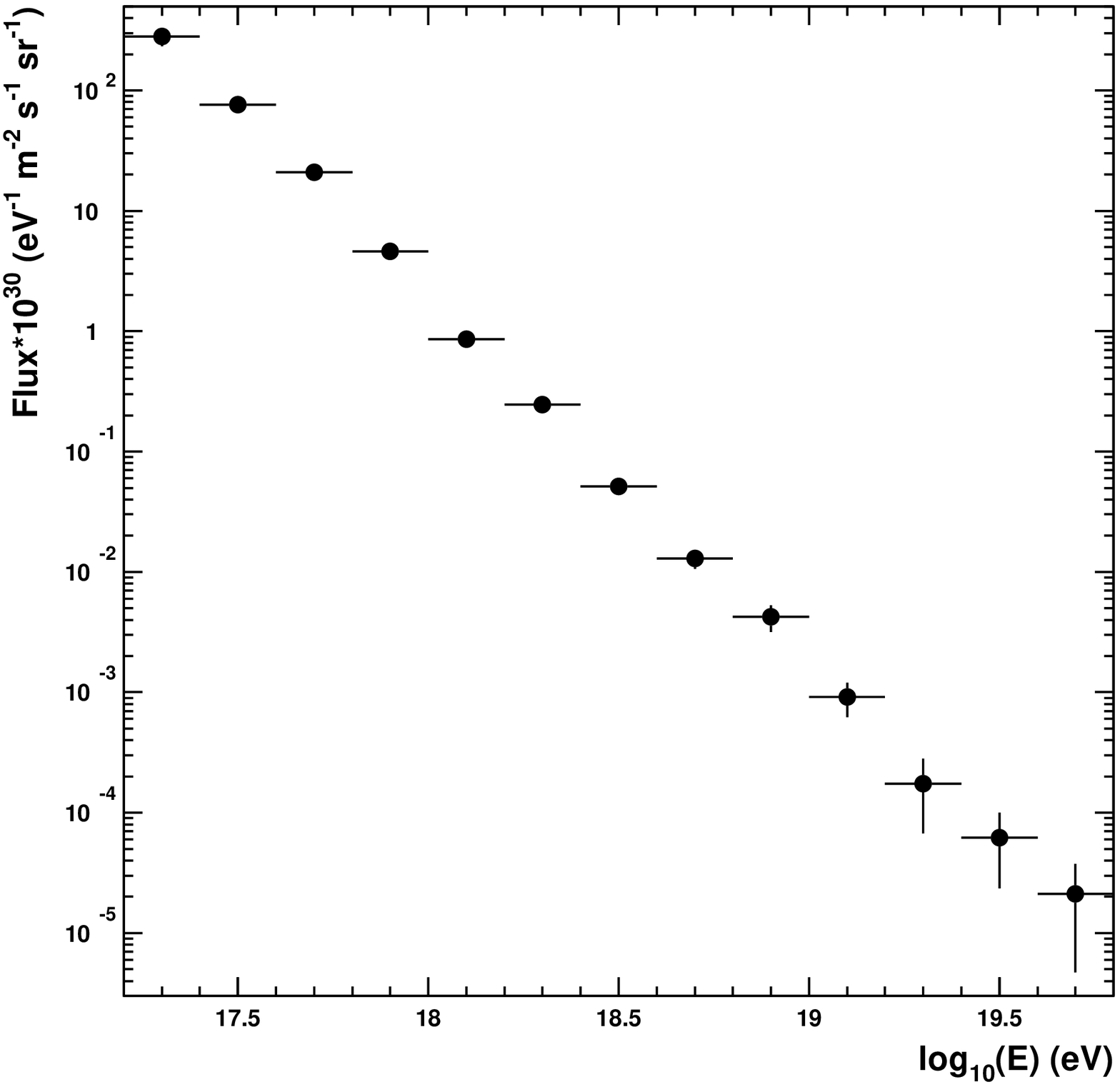}
  \caption{Flux of UHE Cosmic Rays measured by the HiRes-II detector.}
  \label{fig:flux}
\end{figure}

\begin{figure}[tbh]
  \includegraphics[width=\columnwidth]{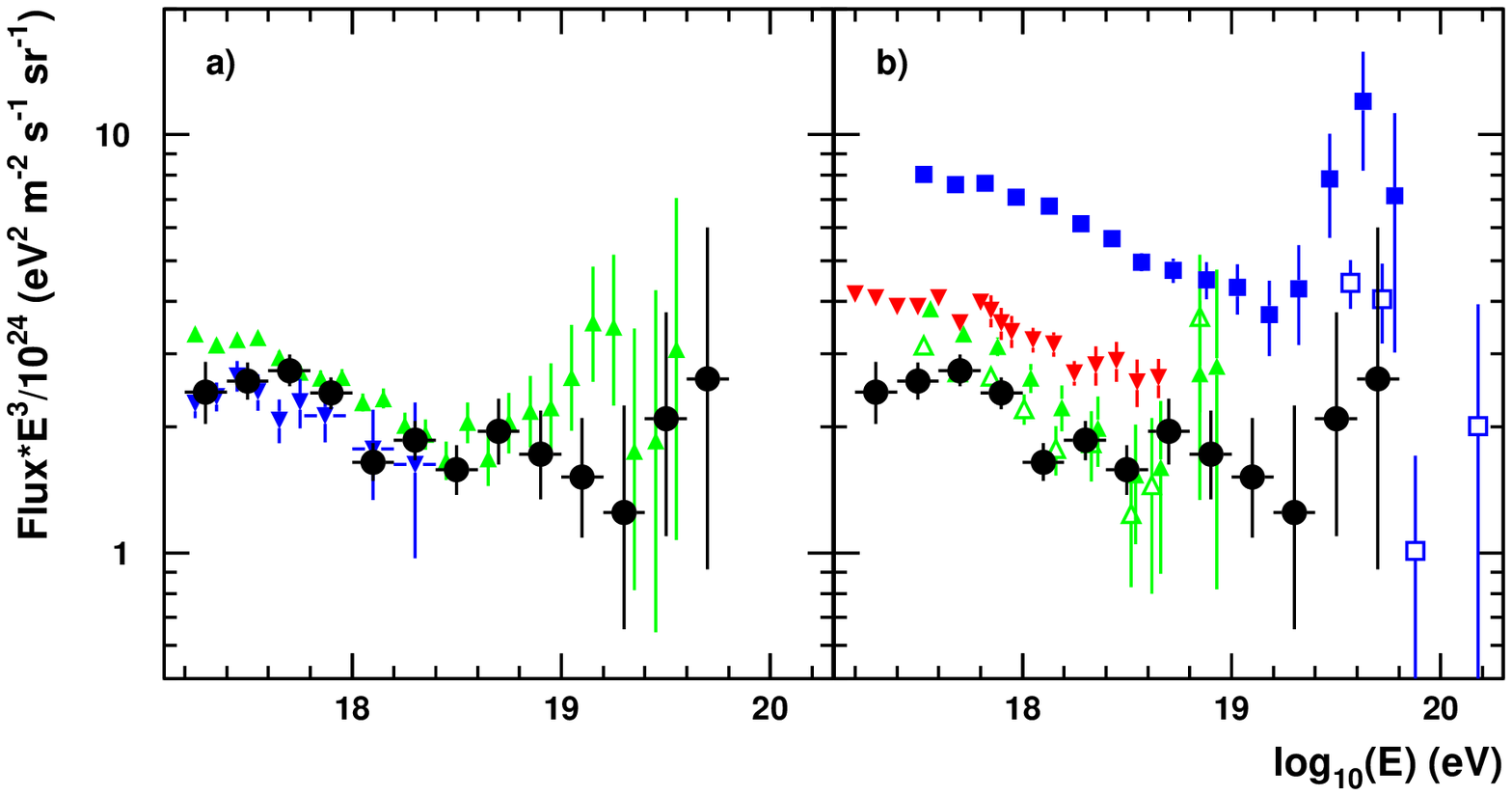}
  \caption{E$^3$ times the HiRes-II UHE Cosmic Ray Flux (filled
    circles), focusing on the energy region just below the ankle.
    Panel a) includes the results of other fluorescence experiments:
    Fly's Eye Stereo\cite{kn:flyseye} (up triangles) and
    HiRes/MIA\cite{kn:hiresmia} (down triangles).  Panel b) includes
    the results of various ground array experiments:
    Akeno\cite{kn:akeno} (down triangles), Haverah Park\cite{kn:hpark}
    proton analysis (filled up triangles) and iron analysis (open up
    triangles), and Yakutsk\cite{kn:yakutsk} trigger-500 (closed
    squares) and trigger-1000 data (open squares).}
  \label{fig:e3flux}
\end{figure}

The measured spectrum, $J(E)$, is shown in Figure \ref{fig:flux}.  The
measured spectrum multiplied by $E^3$ is shown in Figure
\ref{fig:e3flux}.  For the latter, the average energy of the data
events in each bin is used to compute the $E^3$ factor.

Panel a of Figure~\ref{fig:e3flux}) shows the HiRes-II spectrum in
comparison with two previous fluorescence experiments, Fly's
Eye\cite{kn:flyseye} (stereo) and HiRes-MIA\cite{kn:hiresmia}.  The
agreement between the three is quite good.  Since different methods
were used to calibrate the three experiments, one expects slightly
different results.  The three results are all within the calibration
uncertainties of each experiment.  Panel b of Figure~\ref{fig:e3flux}
shows the HiRes-II spectrum in comparison with three ground array
experiments, Akeno\cite{kn:akeno}, Haverah Park\cite{kn:hpark} and
Yakutsk\cite{kn:yakutsk}.  Differences in energy scale calibration
between experiments are accentuated by the $E^3$ factor.

The Fly's Eye experiment, in their stereo analysis, observed the ankle
feature at $3 \times 10^{18}$ eV.  To test whether this feature is
seen in the HiRes-II data, we fit the HiRes-II spectrum to both a
single power law and to a double power law with a floating break
point.  The single power law fit results in a spectra index,
$\gamma=-3.12\pm0.04$, with a $\chi^2= 14.1$ for 13 degrees of
freedom.  The double power law fit results in a spectral index,
$\gamma_1=-3.16\pm 0.05$ below the break point, $\log_{10}E=18.5\pm
0.4$, and a spectral index, $\gamma_2=3.0\pm 0.2$, above the break
point.  The $\chi^2$ for this fit was 12.0 for 11 degrees of freedom.
The $\chi^2$ was reduced by 2.1 while adding 2 parameters.  Since the
$\chi^2$ did not improve significantly, we cannot claim evidence for
the ankle in the HiRes-II monocular data alone.

\section{HiRes-I Analysis}

In addition to the monocular data collected by the HiRes-II detector,
we have a considerable amount of monocular data collected by HiRes-I.
In this section we describe the differences between the two detectors
and their analyses, and in the next section present both monocular
spectra.  For a more complete description of the HiRes-I detector and
its analysis see references\cite{kn:HiRes-INIM}
and\cite{kn:HiRes-IAPP}.

The most important differences between the HiRes-I and HiRes-II
detectors are the time resolution and the number of mirrors.  The
HiRes-II time resolution is about a factor of two better than that of
HiRes-I, and the one ring of mirrors at HiRes-I means that the tracks
are shorter.  These factors affect the resolution of the time vs.
angle plot (a time plot for HiRes-II is shown is the lower left
quadrant of Figure~\ref{fig:event}).  A third difference between the
two detectors' data is that in this paper we are reporting data
covering four years of running for HiRes-I (from 29~May~1997 to
7~Feb.~2003) and six months for HiRes-II.

In reconstructing the geometry of tracks seen by the HiRes-I detector,
we wish to measure $R_p$ and $\psi$ from the curvature in the time
plot (a HiRes-II time plot is displayed in the lower left quadrant of
Figure \ref{fig:event}).  But the shorter tracks means the uncertainty
in $R_p$ and $\psi$ are greater than we would wish for many events.
To solve this problem, we add to our fitting procedure a constraint
based on the longitudinal energy deposition profile of the event (for
a HiRes-II longitudinal profile plot see the lower right quadrant of
Figure \ref{fig:event}).  From previous experiments using fluorescence
detectors\cite{kn:caozh}, and from the HiRes-II analysis reported
here, we know that the Gaisser-Hillas formula in Equation \ref{eq:GH}
fits our events very well.  While $X_{max}$ varies from event to event
and depends logarithmically on the atomic weight of the nucleus and
initial energy, the shape of the shower is largely independent of
these.  Therefore, we use the fact that the shower width does not
change with energy or composition to constrain the fit.

The profile-constrained geometry fit proceeds by first calculating a
combined $\chi^2$ for the time and profile fits.  Each phototube on
the track makes one contribution to the time fit and one to the
profile fit.  A map of $\chi^2$ is made in six steps in $X_{max}$ and
180 steps in $\psi$.  The $X_{max}$ values used are 685, 720, 755,
790, 825, and 960 g/cm$^2$.  These values span the range of $X_{max}$
values expected for our energy range.  The $\psi$ values range from 1
to 180 degrees.  For each of the map points, the fit is performed with
the Gaisser-Hillas parameter $X_0$ fixed to -60 g/cm$^2$.  In the
vicinity of the minimum of the $\chi^2$ map a finer search is
performed, which includes varying the orientation of the
shower-detector plane within bounds of the fired photomultiplier tube
apertures.  To ensure that the reconstruction process has been
accurate, we demand that:
\begin{itemize}
\item The \v{C}erenkov light contribution to the observed flux be less
  than 20\%,
\item The track length be greater than 7.9 degrees,
\item The depth of the first observed point be less than 1000
  g/cm$^2$, 
\item Angular speed $< 3.4^\circ\ \mu s^{-1}$,
\item The average effective mirror area seen by the hit tubes for the
  event $> 0.9$ m$^2$,
\item $\psi < 120^\circ$.
\end{itemize}

\begin{figure}[tbh]
  \includegraphics[width=0.67\columnwidth]{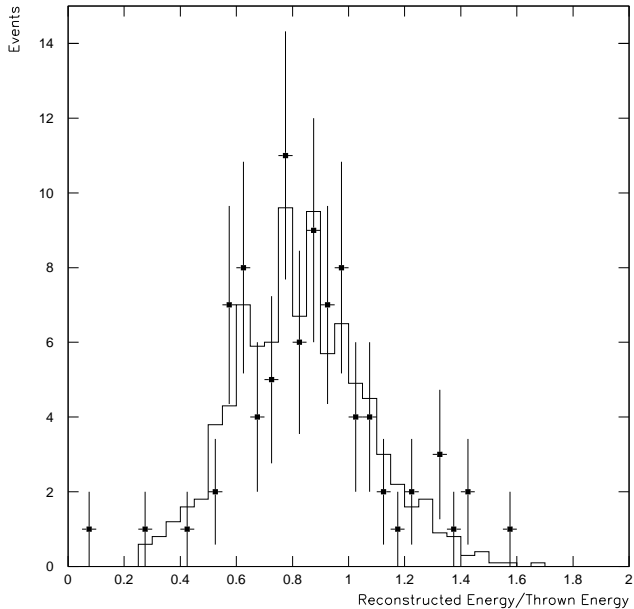}
  \caption{Ratio of HiRes-I reconstructed energy to thrown energy.
    The histogram is for Monte Carlo events.  The black points show
    stereo events from the data where the Monte Carlo thrown energy
    has been replaced with the energy calculated from the stereo
    geometric reconstruction.}
  \label{fig:pcgf}
\end{figure}

\begin{figure}
  \includegraphics[width=0.67\columnwidth]{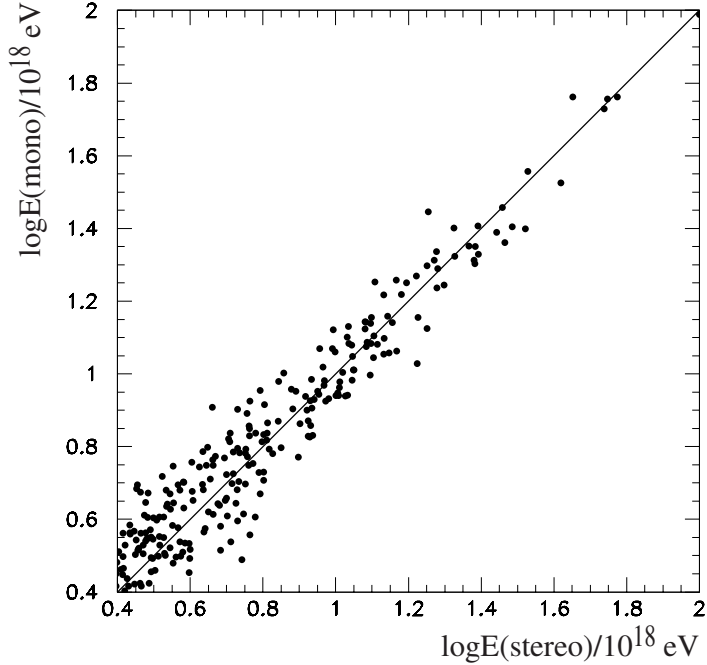}
  \caption{Scatter-plot of energy reconstructed by HiRes-I
    using the profile constrained geometry fit versus the energy
    reconstructed by HiRes-I using the stereo geometry for a set of
    stereo events.}
  \label{hr1_escat}
\end{figure}

In a Monte Carlo study of the profile-constrained geometry fit, we
find that the method works well.  However, it introduces a small bias
into the reconstructed energy.  The bias is 15\% at $3 \times 10^{18}$
eV and falls to 5\% at $3 \times 10^{19}$ eV.  Figure \ref{fig:pcgf}
shows the reconstructed energy divided by the Monte Carlo thrown
energy at $3 \times 10^{18}$ eV.  The bias is evident from the fact
that the peak does not occur at 1.  Superimposed upon this plot is a
similar plot determined from our stereo data.  Here the stereo
information was used to precisely determine the geometry of the event,
but the energy was reconstructed using only information from HiRes-I.
Stereo geometry is like the Monte Carlo in that in both cases the
geometry is well known.  Thus, it is a good test of geometric effects
in reconstruction.  The two curves agree very well.  The shift was
parameterized and a correction applied to the data.  In
Figure~\ref{hr1_escat}, the corrected energy from the
profile-constrained geometry fit is compared to the energy calculated
using stereo geometry.

\begin{figure}[tbh]
  \includegraphics[width=0.67\columnwidth]{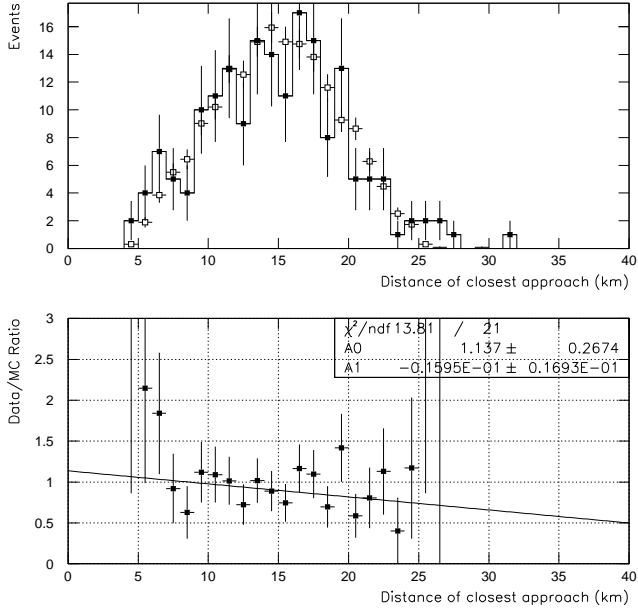}
  \caption{Comparison between data and Monte Carlo for $R_p$, the
    impact parameter of showers, for events where $18.4 < \log{E(eV)}
    < 18.6$}
  \label{fig:mcrp}
\end{figure}

Our Monte Carlo describes the HiRes-I data well.  As an example,
Figure \ref{fig:mcrp} is a comparison between data and Monte Carlo of
$R_p$, the impact parameter of showers, for events where $18.4 <
\log{E(eV)} < 18.6$.  The agreement is excellent.

\section{Systematic Uncertainties}

The largest sources of systematic uncertainty in this experiment are
atmospheric modeling, the absolute calibration of the detector in
units of photons, the absolute yield of the fluorescence process, and
the correction for unobserved energy in the shower.

To test the sensitivity of the flux measurement at HiRes-II to
uncertainties in atmospheric conditions we reanalyzed the data and
generated new Monte Carlo samples with new conditions: we first
changed the aerosol horizontal extinction length from 22 to 20 km,
then we changed the aerosol scale height from 1.1 km to 0.7 km.  The
extinction length change corresponds to one standard deviation.  For
the scale height change we used the RMS of the scale height
distribution, and thus made a conservative estimate of the systematic
uncertainty from this source.  These two variables are related since
the aerosol column depth is equal to their product (for an exponential
atmospheric model).  Changing the horizontal extinction length had
little effect, raising the normalization of $J(E)$ by $(4 \pm 6)$\%.
The change in aerosol scale height had a larger effect, lowering
$J(E)$ on average by $(15 \pm 5)$\%.  We also raised the scale height
and found a symmetric change in $J(E)$.

The systematic uncertainties in the HiRes-I data from atmospheric
conditions are similar to those for HiRes-II.  We found the
reconstructed geometries of HiRes-I events above $10^{18.5}$~eV to be
insensitive to changes in either the aerosol extinction length or the
aerosol scale height, and we saw a maximum change in the energy of
$\pm{13}$\% at $10^{20}$~eV, decreasing to $\pm{6}$\% at
$10^{18.5}$~eV.  Taking the average energy shift, 9\%, the systematic
uncertainty in flux from atmospheric effects at HiRes-I becomes
$\pm{15}$\%.

The systematic uncertainty from the absolute calibration of the
detector is equal to 10\% and is independent of
energy\cite{kn:abscalib}.  The absolute uncertainty in the
fluorescence yield is 10\% and is independent of
energy\cite{kn:kakimoto}.  The uncertainty in the correction for
unobserved energy in the shower is 5\%\cite{kn:chiwha}.  Adding the
uncertainties in quadrature yields a net systematic uncertainty on
$J(E)$, averaged over energy, of 31\%.

\section{Discussion}

\begin{figure}[tbh]
  \includegraphics[width=\columnwidth]{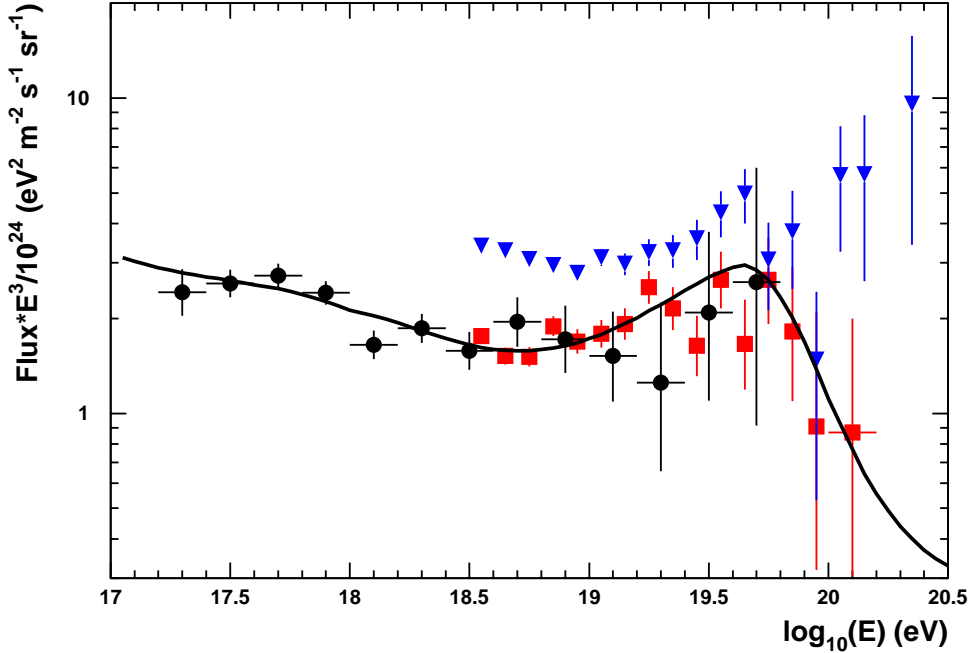}
  \caption{E$^3$ times the UHE Cosmic Ray Flux.  Results from the
    HiRes-I and HiRes-II detectors, and the AGASA experiment are
    shown.  Also shown is a fit to the data assuming a model,
    described in the text, of galactic and extragalactic sources.}
  \label{fig:e3fluxall}
\end{figure}

In Figure \ref{fig:e3fluxall}, the monocular spectra from both the
HiRes-I and HiRes-II detectors are shown\cite{kn:monoPRL}.  In the
energy range where both detectors' data have good statistical power
the results agree with each other very well.  The highest energy
HiRes-I data point corresponds to two events reconstructed at $1.0$
and $1.5\times 10^{20}$ eV.  

We now fit the combined HiRes-I and HiRes-II monocular spectra to both
a single power law fit and a double power law fit with a floating
break point.  The single power law fit results in a spectra index,
$\gamma=-3.07\pm0.02$, with a $\chi^2= 67.8$ for 31 degrees of
freedom.  This is not an acceptable fit.  The double power law fit
results in a spectral index, $\gamma_1=-3.17\pm 0.03$ below the break
point, $\log_{10}E=18.65\pm 0.05$, and a spectral index,
$\gamma_2=2.89\pm 0.04$, above the break point.  The $\chi^2$ for this
fit was 41.1 for 29 degrees of freedom.  The large improvement in the
$\chi^2$ (26.7 while adding only two parameters) indicates strong
evidence for the ankle being present in the combined HiRes monocular
data.

The latest results of the AGASA experiment are also shown in this
figure\cite{kn:agasa}.  Below about $1 \times 10^{20}$ eV the AGASA
results are consistently a factor of two higher than ours.  Above this
energy their data points diverge from the trend of our data.  Since
the vertical axis in Figure \ref{fig:e3fluxall} is $E^3 \times J(E)$ a
modest change in the energy scale would bring the experiments into
considerably better agreement.  For example lowering the AGASA energy
scale by 30\% would bring their points down by a factor of 2, move
them to the left by 0.15 in $\log(E)$, and reduce the discrepancy
between the two experiments.  Such a change is within the systematic
uncertainties of each experiment.

In the energy range, $18.7 < \log{E} < 19.8$ the HiRes data is fit by
an $E^{-2.8}$ power law.  The three highest-energy data points do not
lie along an extension of that power law.  Such an extension would
predict that 25.3 events would occur above $\log{E} = 19.8$ while only
10 were seen.  The Poisson statistics probability of observing 10 or
fewer events while expecting 25.3 is $4.9 \times 10^{-4}$.  

On the other hand, our data are consistent with the prediction of a
GZK cutoff.  As an example of what one would expect we have fit the
data to a model that consists of two sources for cosmic rays, galactic
and extragalactic, which includes the GZK threshold\cite{kn:bergman}.
We use the extragalactic propagation model of Berezinsky, Gazizov, and
Grigorieva\cite{kn:berezinsky}, modified to take account of discrete
energy losses of protons as in the paper by Blanton, Blasi and
Olinto\cite{kn:blanton}, and assume that protons come from sources
distributed uniformly following the expansion of the universe, and
lose energy by pion and $e^+ e^-$ production from the cosmic microwave
background radiation, as well as from the expansion of the universe.
Since the measured composition\cite{hrmcomp,hrscomp} changes from
heavy to light within our energy range, we approximate the galactic
component of cosmic rays as being the fraction of iron.  We take this
fraction to be 55\% at $10^{17}$ eV, decreasing linearly with
$\log{E}$ to 20\% at $10^{17}$ eV, then decreasing to zero at
$10^{20}$ eV.  The model includes an end to the extragalactic input
spectrum at $1 \times 10^{21}$ eV.  The fitting parameters of the
model are the normalization and power law index (at the source) of
extragalactic cosmic rays.  The power law index in the fit was -2.4.
The fit is excellent with $\chi^2$ of 32.6 for 31 degrees of freedom.
In this model, the peak at $\log{E}$ of 19.8 is due to fitted
$E^{-2.4}$ input spectrum being cut off at the pion production
threshold, the ankle is due to energy losses from $e^+ e^-$
production, and the second knee comes from the $e^+ e^-$ production
threshold.

\section{Conclusions}

We have measured the flux of UHE cosmic rays with the FADC detector of
the HiRes experiment.  Use of Flash ADC information allowed us to
reduce systematic errors in reconstruction of events.  We developed
our Monte Carlo simulation programs to very accurately model the
experiment, and calculated the exposure of the experiment in a way
that takes into account the experimental resolution.  The result
reported here is in good agreement with the cosmic ray flux
measurement made with the HiRes-I detector.  The latter measurement is
based on a largely statistically independent data set, with only a
limited number of stereo events in common to both analyses.  The
result reported here is also consistent with the flux measured by the
Fly's Eye experiment using the stereo reconstruction technique.  Above
$10^{20}$eV our data is significantly different from that of the AGASA
experiment.  The ankle is not seen in the HiRes-II monocular alone,
but is apparent in the combined HiRes-I and HiRes-II data.  We have
fit our data to a model incorporating both galactic and extragalactic
sources of cosmic rays, which includes the GZK cutoff, and find good
agreement.

\section{Acknowledgements}

This work is supported by US NSF grants PHY-9321949, PHY-9322298,
PHY-0098826, PHY-0245428, PHY-0305516, PHY-0307098, by the DOE grant
FG03-92ER40732, and by the Australian Research Council.  We gratefully
acknowledge the contributions from the technical staffs of our home
institutions.  The cooperation of Colonels E. Fischer and G. Harter,
the US Army, and the Dugway Proving Ground staff is appreciated.


\begin{thebibliography}{99}
  
\bibitem{kn:acceleration} R.J. Protheroe, Topics in Cosmic Ray
  Astrophysics (ed. M.A. DuVernois, Nova Science Publishing, NY, 1999)
  and astro-ph/9812055.
  
\bibitem{kn:gzk} K. Greisen, Phys. Rev. Lett. {\bf 16}, 748 (1966); G.
  T. Zatsepin and V. A. Kuzmin, Pis'ma Zh. Eksp. Teor. Fiz. {\bf 4},
  114 (1966) [JETP Lett. {\bf 4}, 78 (1966)].
  
\bibitem{kn:prevexp} J. Linsley, Phys. Rev. Lett. {\bf 10}, 146 (1963)
  and Proc. 8th Int. Cosmic Ray Conf. {\bf 4}, 295 (1963).\\
  M.A. Lawrence, R.J.O. Reid, and A.A. Watson, J. Phys. G
  Nucl. Part. Phys. {\bf 17}, 733 (1991) and references therein.\\
  B.N. Afanasiev {\it et al.}, Proc. Tokyo Workshop on Techniques for
  the Study of Extremely High Energy Cosmic Rays (ed. M. Nagano, Inst.
  for Cosmic Ray Research, Univ. of Tokyo), 35 (1993).

\bibitem{kn:flyseye} D. J. Bird {\it et al.}, Phys. Rev.  Lett. {\bf
    71}, 3401 (1993) and Astrophys. J. {\bf 441}, 144 (1995).  See
  also T. Abu-Zayyad {\it et al.}, Astrophys. J. {\bf 557}, 686
  (2001).
  
\bibitem{kn:agasa} M~Takeda {\it et al.}, Astropart.~Phys. {\bf 19},
  447 (2003).
  
\bibitem{kn:knee} A.A. Watson, Proc. 25th Int. Cosmic Ray Conf.
  (Durban), {\bf 8}, 257 (1997) and references therein.
  
\bibitem{kn:akeno} M. Nagano {\it et al.}, J. Phys. G. {\bf 10}, 1295
  (1984) and M. Nagano {\it et al.}, J. Phys. G. {\bf 17}, 733 (1991).
  
\bibitem{kn:hpark} M. Ave {\it et al.}, Proc. 27th Int. Cosmic Ray
  Conf. (Hamburg) {\bf 1}, 381 (2001).  See also astro-ph/0112253.
  
\bibitem{kn:yakutsk} M.I. Pravdin {\it et al.}, Proc. 26th Int. Cosmic
  Ray Conf. (Salt Lake City), {\bf 3}, 292 (1999) and M.I. Pravdin
  {\it et al.}, Proc. 28th Int. Cosmic Ray Conf. (Tuskuba), 389
  (2003).
  
\bibitem{kn:HiRes-INIM} T. Abu-Zayyad {\it et al.}, Proc. 26th Int.
  Cosmic Ray Conf. (Salt Lake City), {\bf 5}, 349 (1999).
  
\bibitem{kn:HiRes-IINIM} J. Boyer, B. Knapp, E. Mannel, and M. Seman,
  Nucl. Instr. Meth. {\bf A482}, 457 (2002).
  
\bibitem{kn:abscalib} R.U.~Abassi {\it et al.}, to be submitted to
  Nucl. Instr. Meth. 
  
\bibitem{kn:avgatm} R.U.~Abassi {\it et al.}, in preparation, and
  http://www.cosmic-ray.org/atmos/.
  
\bibitem{kn:balt} R. Baltrusaitis {\it et al.}, Nucl. Inst. and
  Meth. {\bf A240}, 410 (1985).  See also P. Sokolsky, ``Introduction
  to Ultrahigh Energy Cosmic Ray Physics,'' (Cambridge University
  Press, 1990).
  
\bibitem{kn:bunner} A. N. Bunner, Ph.D. Thesis, Cornell University,
  Ithaca, NY (1964).
  
\bibitem{kn:kakimoto} F. Kakimoto {\it et al.}, Nucl. Instr. Meth.
  {\bf A372}, 527 (1996).  See also G. Davidson and R. O'Neil, J.
  Chem. Phys. {\bf 41}, 3946 (1964).
  
\bibitem{kn:ghfit} T. K. Gaisser and A. M. Hillas, Proc. 15th Int.
  Cosmic Ray Conf. (Plovdiv), {\bf 8} 353, (1977).
 
\bibitem{kn:caozh} T. Abu-Zayyad {\it et al.}, Astropart. Phys.  {\bf
    16}, 1 (2001).
  
\bibitem{kn:chiwha} C. Song {\it et al.}, Astropart. Phys.  {\bf 14},
  7 (2000).  See also J. Linsley, Proc. 18th ICRC, Bangalore, India,
  {\bf 12}, 144 (1983) and R.M. Baltrusaitis {\it et al.}, Proc. 19th
  ICRC, La Jolla, USA, {\bf 7}, 159 (1985).

\bibitem{kn:corsika} Heck, D., Knapp, J., Capdevielle, J. N., Schatz,
  G., Thouw, T., Report FZKA 6019 (1998), Forschungszentrum
  Karls\-ruhe; http://www-ik3.fzk.de/
  heck/corsika/physics\_description/ corsika\_phys.html
 
\bibitem{kn:qgsjet} N. N. Kalmykov, S. S. Ostapchenko, A. I. Pavlov,
  Nucl. Phys. B (Proc. Suppl.) {\bf 52B}, 17 (1997).

\bibitem{cowan} G.~Cowan, ``Statistical Data Analysis,'' Oxford
  Univ. Press, NY, (1998); see Chapter 11.
  
\bibitem{kn:hiresmia} T.~Abu-Zayyad {\it et al.}, Astrophys.~J. {\bf
    557}, 686 (2001).

\bibitem{kn:HiRes-IAPP} R.U.~Abassi {\it et al.}, to be submitted to
  Astropart. Phys.
  
\bibitem{kn:monoPRL} R.U.~Abassi {\it et al.}, accepted by
  Phys.~Rev.~Lett., astro-ph/0208243.
  
\bibitem{kn:bergman} See R.U.~Abassi {\it et al.}, in preparation.
  See also E.~Waxman, Astrophys.~J.~Lett. {\bf 452}, L1 (1995); and
  J.N.~Bahcall and E.~Waxman, Phys.~Lett. {\bf B556}, 1 (2003).
  
\bibitem{kn:berezinsky} V. Berezinsky, A.Z. Gazizov, S.I Grigorieva,
  hep-ph/0204357.  See also S.T.~Scully and F.W.~Stecker,
  Astropart.~Phys. {\bf 16}, 271 (2002).

\bibitem{kn:blanton} M.~Blanton, P.~Blasi and A.V.~Olinto,
  Astropart.~Phys. {\bf 15}, 275 (2001).

\bibitem{hrmcomp} T.~Abu-Zayyad {\it et al.}, Phys. Rev. Lett. {\bf
    84}, 4276, (2000).

\bibitem{hrscomp} R. Abbasi {\it et al.}, submitted to Ap. J.,
  astro-ph/0407622.

\end{thebibliography}
\end{document}